# Monolayer 1T-NbSe$_2$ as a 2D correlated magnetic insulator

Mengke Liu[1], Joshua Leveillee[1,2], Shuangzan Lu[3], Jia Yu[1], Hyunsue Kim[1], Keji Lai[1], Chendong Zhang[3], Feliciano Giustino[1,2], Chih-Kang Shih[1]*

[1] Department of Physics, The University of Texas at Austin, Austin, TX 78712, USA

[2] Oden Institute for Computational Engineering and Sciences, The University of Texas at Austin, Austin, TX 78712, USA

[3] School of Physics and Technology and Key Laboratory of Artificial Micro- and Nano-structures of Ministry of Education, Wuhan University, Wuhan 430027, China

*email: shih@physics.utexas.edu.

**Abstract**

**Monolayer group-V transition metal dichalcogenides in their 1T phase have recently emerged as a platform to investigate rich phases of matter, such as spin liquid and ferromagnetism, resulting from strong electron correlations. Although 1T phase NbSe$_2$ does not occur naturally in bulk form, it has been discovered that the 1T and 1H phases can coexist when monolayer NbSe$_2$ is grown via molecular beam epitaxy (MBE). This discovery has inspired theoretical investigations predicting collective phenomena such as ferromagnetism in two dimensions. Here, by controlling the MBE growth parameters, we demonstrate the successful growth of single-phase 1T-NbSe$_2$. By combining scanning tunneling microscopy/spectroscopy and *ab initio* calculations, we show that this system is a charge-transfer insulator, with the upper Hubbard band located above the valence band maximum. Furthermore, by creating a vertical 1T/2H NbSe$_2$ heterostructure, we find evidence of exchange interactions between the localized magnetic moments in 1T phase and the metallic/superconducting phase, as manifested by Kondo resonances and Yu-Shiba-Rusinov bound states.**

**Introduction**

Strong electron correlations underpin many unique phases of quantum matter, including unconventional superconductivity (*1, 2*), correlated insulator (*3-6*), magnetism (*3, 7*), spin liquids (*8*), and Wigner crystals (*6, 9, 10*). Recently, group-V transition metal dichalcogenides (MX2, with M = V, Nb, Ta; and X = S, Se) have emerged as an interesting platform to explore strong electron correlations (*11-16*). Compared to their group-VI TMDs (M = Mo, W) neighbors in the periodic table, which are semiconducting, group-V TMDs have one less valence electron, which makes them nominally metallic in the 2H phase (*17*). Many group-V TMDs also exhibit a 1T phase. In this phase, TMDs undergo a charge density wave (CDW) transition at low temperatures (*17-19*). It is also believed strong electron correlations manifest in this CDW phase and give rise to exotic magnetic properties, especially in the monolayer regime. For example, monolayer 1T-$VSe_2$ is reported to have room temperature ferromagnetism (*20*), which has stimulated an intense debate both theoretically (*21*) and experimentally (*22-24*). Monolayer 1T-$TaSe_2$ is reported to be a Mott insulator without long-range magnetic order (*25*). Monolayer 1T-TaS2 undergoes successive phase transitions as a function of temperature (*26*) and is predicted to be a quantum spin liquid at low temperatures (*27*).

Newly emerging is the 1T phase $NbSe_2$ which does not occur naturally in bulk crystals but was discovered to coexist with the 1H phase during MBE growth of monolayer $NbSe_2$ on graphene (*28*). The existence of 1T-$NbSe_2$ has stimulated many theoretical efforts to explore whether correlated electronic phenomena exist in this phase. The calculations revealed that the correlation gap is of charge-transfer nature and showed that soft phonons are responsible for the David star CDW (*29-31*). Moreover, it has been suggested that the interplay of a CDW and strong electron correlation could lead to the existence of ferromagnetism in monolayer 1T-NbSe$_2$ (*29, 30*). These prior studies indicated that the 1T phase group-V TMDs form a very interesting platform where strong electron correlations are responsible for exotic magnetic phases but yet to be experimentally verified.

Here we report the controlled growth of pure 1T-NbSe$_2$, pure 1H-NbSe$_2$, and their mixed phases in the monolayer regime by fine-tuning the MBE parameters. The electronic structure of monolayer 1T-NbSe$_2$ is probed using scanning tunneling microscopy/spectroscopy (STM/STS). By combining our spectroscopic data with Hubbard-corrected density-functional theory (DFT) calculations, we show that the gap is indeed of charge-transfer nature, with a localized upper Hubbard band (UHB) located above the valence band maximum (VBM). The lower Hubbard band (LHB) is found to merge with the valence band. Detailed orbital textures as a function of energy are mapped out and compare favorably with the theoretical calculations. We also find that the localized Hubbard band can be easily destroyed by point defects. To probe magnetism, we create a vertical heterostructure comprising of monolayer 1T-NbSe$_2$ on top of metallic/superconducting 2H-NbSe$_2$. Our STS measurements on this heterostructure reveal Kondo resonances (*32, 33*) and Yu-Shiba-Rusinov bound states (*34-36*), which are clear fingerprints of the exchange interaction between localized magnetic moments in 1T-NbSe$_2$ and the electrons/Cooper pairs in 2H NbSe$_2$. In addition, the spatial modulation of Kondo peak amplitude is shown to correlate with the calculated spin density across the David star CDW, thus providing direct and unambiguous evidence for the existence of large localized magnetic moments driven by electron correlations.

## Results

**Growth mechanism and electronic structure of monolayer 1T-NbSe$_2$**

Bulk NbSe$_2$ does not occur naturally in the 1T polymorph. Using MBE, we can achieve controlled growth of the 1T and 1H polymorphs of monolayer NbSe$_2$ (Fig. 1A and B) on both graphite and graphene substrates by changing the growth temperature and the Se to Nb flux ratio. Fig. 1C shows the parameter space used to control the growth of monolayer NbSe$_2$ on HOPG substrates. Note in this parameter space the same Se deposition rate is kept, while the Nb deposition rate varies. Thus, a high Se to Nb flux ratio can also be interpreted as a low growth rate. Under this condition, the pure 1T phase is achieved with a high Se to Nb flux ratio over a reasonable range of growth temperature. A reflection high-energy electron diffraction pattern taken on monolayer NbSe$_2$ on a bilayer graphene substrate (Fig. 1D) shows the good crystallinity of our sample. While pure 1T-NbSe$_2$ can be obtained on a graphite substrate, as shown in Fig. 1E and supplementary Fig. S1A, the same parameters lead to mixed 1H and 1T phases when using epitaxial graphene substrates (graphene-terminated 6H-SiC (0001) substrates) (Fig. 1F), implying that the parameter space for the growth of 1T-NbSe$_2$ on graphene or graphite differs. Additional discussions can be found in supplementary Fig. S1.

A $\sqrt{13} \times \sqrt{13}$ CDW is observed on 1T-NbSe$_2$ (Fig. 1G), in direct contrast to the $3 \times 3$ CDW in the 1H phase (Fig. 1H). Our atomic-scale image of 1T-NbSe$_2$ (Fig. 2A) reveals that the rotation angle of the charge density wave with respect to its crystal lattice is 13.9°. A 1T phase NbSe$_2$ lattice model stacked on top of the STM image (Fig. 2A) demonstrates the David star shape of the CDW in a supercell consisting of 13 Nb atoms. Within the single-particle picture, this system should be metallic since each David star unit cell contains an odd number of electrons with each Nb atom contributes one d electron. However, our d$I$/d$V$ STS data reveals an insulating ground state (Fig. 2B). In particular, we find a 150 meV gap, and we clearly resolve an isolated peak in the density of states (DOS) at 170 meV. Several pronounced peaks are also seen below the Fermi level, however, embedded in a broad continuum background rather than isolated. This insulating ground state can be described by the charge-transfer insulator scenario (*3*), as schematically illustrated in Fig. 2C. The strong electron correlation opens a charge gap in the original half-filled band. The band pushed above the Fermi level is labeled as UHB, and the one below the Fermi level is labeled as LHB. The LHB is pushed low enough that it hybridizes with the valence band continuum. Note that these observations directly differ from the previous reported 0.4 eV gap in a Mott-Hubbard insulator scenario (*28*).

This interpretation is consistent with our DFT+U calculations. Shown in Fig. 2D is the band structure calculated with U = 2.95 eV. In the calculation, the energy is referenced to the valence band maximum, with the upper Hubbard band (UHB) and lower Hubbard band (LHB) highlighted in red and their spins in the opposite directions. The UHB is nearly isolated and flat, with a bandwidth of only 25 meV, while the LHB falls within the valence band continuum, thus demonstrating the charge-transfer character of this system. In addition, a few pronounced spectral peak features seen in the d$I$/d$V$ spectrum (Fig. 2B) are also captured in the calculated DOS (Fig. 2E). A careful comparison between calculations and experiments allows us to identify the Γ point energy of the UHB and LHB, as indicated Fig. 2B and Fig. 2E. Although the calculated band gap is slightly larger than in the experiments (by ~ 0.1 eV), the calculated DOS shows good agreement with the experimental spectrum. The fact that UHB and LHB are spin-polarized and well separated from the Fermi level suggests the existence of a net magnetic moment and potential magnetic order. We will come back to this point when discussing Fig. 5.

**Orbital texture mapping of 1T-NbSe$_2$**

We further probe the energy-dependent orbital textures and compare the results with our calculations. We use a constant-height mode to map the d$I$/d$V$ image (Fig. 3A to E). Previously this mode has been shown to reflect the orbital texture better than the constant current d$I$/d$V$ image (*37, 38*). The calculated charge density distribution is superimposed to the STM image (Fig. 3A to E) and is consistent with the experimental map. Among the valence band, the orbital textures show similar features; thus, we only show two energies here (Fig. 3A and B). The d$I$/d$V$ images follow the symmetry of the David star CDW, with most charge density concentrated on the center of each star (Fig. 3A and B). A richer orbital texture is found in the conduction band. The orbital texture corresponding to the UHB energy (Fig. 3C) is strongly localized around the central Nb atom. This localization is consistent with the very small bandwidth of the UHB (Fig. 2D). At higher energies, the orbital texture appears to be smeared out (Fig. 3D) and it gradually evolves to become delocalized, forming mesh structures and connecting with other David stars (Fig. 3E and F), in line with the dispersive nature expected of states in the conduction band continuum.

**The fragility of UHB at local defect positions**

To gain more insight into the properties of the Hubbard band and its localized charge distribution, we investigate the impact of point defects. Shown in Fig. 4A are two defects most commonly observed in 1T-NbSe$_2$. We label such defects as type I and type II based on the morphology of their associated CDW. The type I defect appears as a weakened CDW spot, while the type II defect appears as a missing CDW spot. In direct contrast to the d$I$/d$V$ spectra on pristine 1T-NbSe$_2$ (Fig. 4B, Fig. 2B), the d$I$/d$V$ spectra on type I and type II defects (Fig. 4C, D) miss the UHB feature (See supplementary Fig. S2A for more details). As a result, the insulating gap at these defects' positions is much wider, reaching up to about 0.45 eV. As previously reported, the most common defects in TMDs are either a metal atom vacancy or a chalcogen atom vacancy (*39*). From the discussion in Fig. 2 and Fig. 3, we know that the UHB is strongly localized on the central Nb atom of the David star in pristine 1T-NbSe$_2$; therefore, we hypothesize that a missing David star feature may be associate with a missing central Nb atom. We theoretically choose a Nb atom vacancy model (see supplementary Fig. S2B) to simulate type II defects. The calculated DOS of this model (Fig. 4E) indeed shows the disappearance of the UHB, consistent with our experimental observation. While a more thorough exploration of point defects in 1T-NbSe$_2$ is warranted, this model confirms the crucial role of the central Nb atom. Putting together the information gathered so far on pristine and defective 1T-NbSe$_2$, we infer that the David star is crucially linked with the spin-polarized electron density around the central Nb atom and is very sensitive to the local bonding environment.

**Localized magnetic moment**

To explore the magnetic properties of 1T-NbSe$_2$, we create a 1T-NbSe$_2$/2H-NbSe$_2$ vertical heterostructures following the approach of Bischoff et al. (*40*), as shown in Fig. 5A, B and supplementary Fig. S3A, B. If the magnetic moments exist in the 1T phase, as predicted by our calculations, their interactions with the electrons near the Fermi surface of the 2H phase should lead to Kondo resonances. Indeed, this is what the experiment shows. Fig. 5C shows the spatial d$I$/d$V$ mapping acquired at 0.35 K and Fig. 5D shows the corresponding spin density distribution, majority spin minus minority spin, calculated by DFT. The d$I$/d$V$ mapping (Fig. 5C) and the selected spectra (Fig. 5E) show clear Kondo resonances; in addition, the spatial modulation of the Kondo resonance peak amplitude (Fig. 5C, E) correlates very well with the calculated spin density distribution in the 1T phase (Fig. 5D). These data provide unambiguous evidence that the observed Kondo resonances result from the interaction between Fermi surface electrons in the 2H phase (metal) and the magnetic moment of the David star in the 1T phase (charge-transfer insulator).

Since bulk 2H-NbSe$_2$ is also a superconductor with a critical temperature of 7.2 K, the exchange interaction between the magnetic moments in the 1T phase and the Cooper pairs in the 2H phase should also lead to Yu-Shiba-Rusinov bound states below the superconducting critical temperature (*34-36*). The spectra shown in Fig. 5C and E are acquired with a lock-in amplifier modulation amplitude of 2.0 meV (peak-to-valley), exceeding the bound states energy separation. With a finer energy resolution (0.1 meV energy step and 0.4 meV lock-in modulation amplitude), the bound states are indeed observed. In Fig. 5F, we show two representative spectra acquired on the 1T phase, along with the superconducting gap observed on the neighboring 2H phase. In the 1T region, multiple YSR bound states appear superimposed to the Kondo resonance background, thus demonstrating the interplay between Kondo physics and superconductivity (*41-44*). While the binding energies of these bound states and the electron-hole spectral weight asymmetry depend on the microscopic details of the exchange interaction and superconducting pairing (*42, 45, 46*), and certainly deserve a separate investigation, our current measurements provide additional strong evidence for magnetism in 1T-NbSe$_2$.

**Discussion**

In this work, we have achieved a controlled growth of monolayer single-phase 1T-NbSe$_2$ on both graphite and bilayer graphene/SiC substrates, and we demonstrated that this new system is a charge-transfer insulator. We find a non-dispersive UHB associated with the central Nb atom of the David star resulting from the $\sqrt{13} \times \sqrt{13}$ CDW reconstruction, and we show that this feature is very sensitive to the local bonding environment. By fabricating insulator/superconductor heterostructures of 1T and 2H NbSe$_2$, we provide direct and unambiguous evidence for localized magnetic moments via the observation of Kondo resonances and Yu-Shiba-Rusinov bound states. Further investigations are needed to conclude whether long-range magnetic order exists in this system. Having demonstrated the controlled growth and orbital texture mapping of pure phases of 1T-NbSe$_2$ and 2H-NbSe$_2$ as well as their mixed phases and heterostructures, we believe NbSe$_2$ now offers a uniquely versatile platform to investigate the interplay of charge, spin, and lattice degrees of freedoms in two dimensions, as well as to advance our understanding of magnetism and superconductivity in atomically thin crystals.

**Methods**

**Sample growth and STM/STS measurements**

Monolayer 1T-NbSe2 was grown in a home-built molecular beam epitaxy chamber with base pressure at $\sim 10^{-10}$ torr. High purity Nb (99.9%) and high purity Se (99.999%) were evaporated from an e-beam evaporator and a standard Knudsen cell, respectively. The flux ratio was determined using crystal monitor. Samples were transferred from MBE into STM, base pressure $\sim 10^{-11}$ torr, through a transfer vessel, base pressure $\sim 10^{-10}$ torr, to maintain the perfect crystallinity of the film. STM/STS measurements were conducted at 4.3 K except for the measurements of Kondo and YSR bound states, which were performed at 0.35 K. The W tip was prepared by an electrochemically etched W tip treated with in situ electron-beam cleaning. STM d$I$/d$V$ spectra were measured using a standard lock-in technique with feedback loop off, whose modulation frequency is 490 Hz. Constant height d$I$/d$V$ mappings were performed with feedback loop on.

**Density-functional theory calculations**

Density-functional theory calculations are performed in the Quantum ESPRESSO software package (*47*). The PBE exchange correlation-function is used to approximate the exchange and correlation energy (*48*). Nb and Se scalar relativistic optimally norm-conserving Vanderbilt (ONCV) pseudopotentials from the PseudoDojo repository are utilized (*49*). On-site correlation on the Nb-4d states are treated by performing DFT+U calculations including spin-polarization (*50*). A plane wave energy cutoff of 120 Ry, a charge density energy cutoff of 480 Ry, a Gamma-centered 4 × 4 × 1 k-point integration grid, and 20 Angstroms of vacuum are used to converge the total energy of the cell.

**Acknowledgments**

We are grateful to Di Xiao and Ming Xie for helpful discussions. This work was primarily supported by the National Science Foundation through the Center for Dynamics and Control of Materials: an NSF MRSEC under Cooperative Agreement No. DMR-1720595. J. L. and F. G. were supported by the Robert A. Welch Foundation under Award number F-1990-20190330. Other supports were from NSF Grant Nos. DMR-1808751, DMR-1949701, the Welch Foundation F-1672 and the National Natural Science Foundation of China (Grant No. 11774268, 11974012).

**Author Contributions**

M.L. and C-K.S. designed and coordinated the experiments; M.L. and S.L. carried out the STM/STS measurements; M.L., J.Y., and H.K. performed the sample growth. J.L. performed the DFT calculations. M.L., S.L., C.Z., and C-K.S. analyzed the STM/STS data; J.L. and F.G. performed the theoretical analyses. M.L., F.G., and C-K.S. wrote the paper with input from all authors. All authors contributed to the scientific discussion.

**Competing financial interests**

The authors declare no competing financial interest.

**Data and materials availability**

All data needed to evaluate the conclusions in the paper are present in the paper and/or the Supplementary Materials. Additional data related to this paper may be requested from the authors.

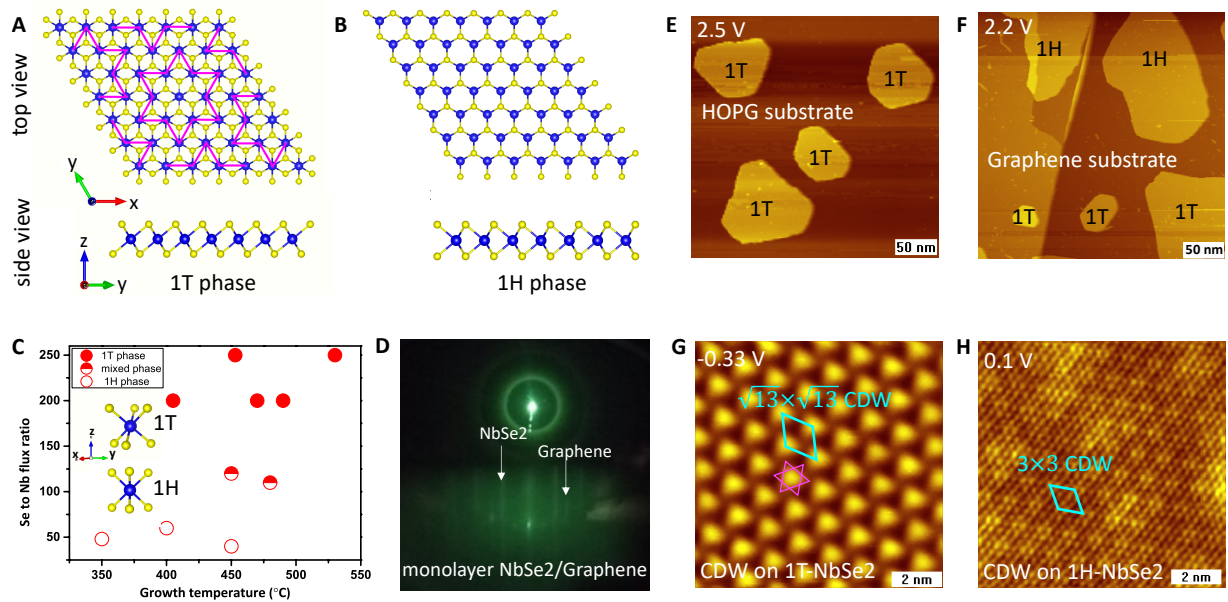

**Fig. 1. Growth parameter and structural characterization of monolayer NbSe$_2$.** (**A** and **B**) Schematic illustrations of 1T and 2H polymorphs of monolayer NbSe$_2$. The Nb (Se) atoms are shown in blue (yellow). The magenta lines in (A) outline the $\sqrt{13} \times \sqrt{13}$ David star CDW in the 1T phase. (**C**) Growth parameter space for 1T and 1H NbSe$_2$ as a function of growth temperature and the Se to Nb flux ratio. Se to Nb flux ratio is varied under constant Se vapor pressure. (**D**) The reflection high-energy electron diffraction pattern of monolayer NbSe$_2$ on bilayer graphene/SiC substrate. (**E** and **F**) Large scale topographic images of NbSe$_2$ flakes. (**E**) shows single 1T phase on HOPG substrate; (F) shows coexisting 1T and 1H phases on graphene substrate. (**G** and **H**) STM images of 1T-NbSe$_2$ and 1H-NbSe$_2$. The $\sqrt{13} \times \sqrt{13}$ David star CDW unit cell is marked in (G); the 3× 3 CDW unit cell is marked in (H**)**.

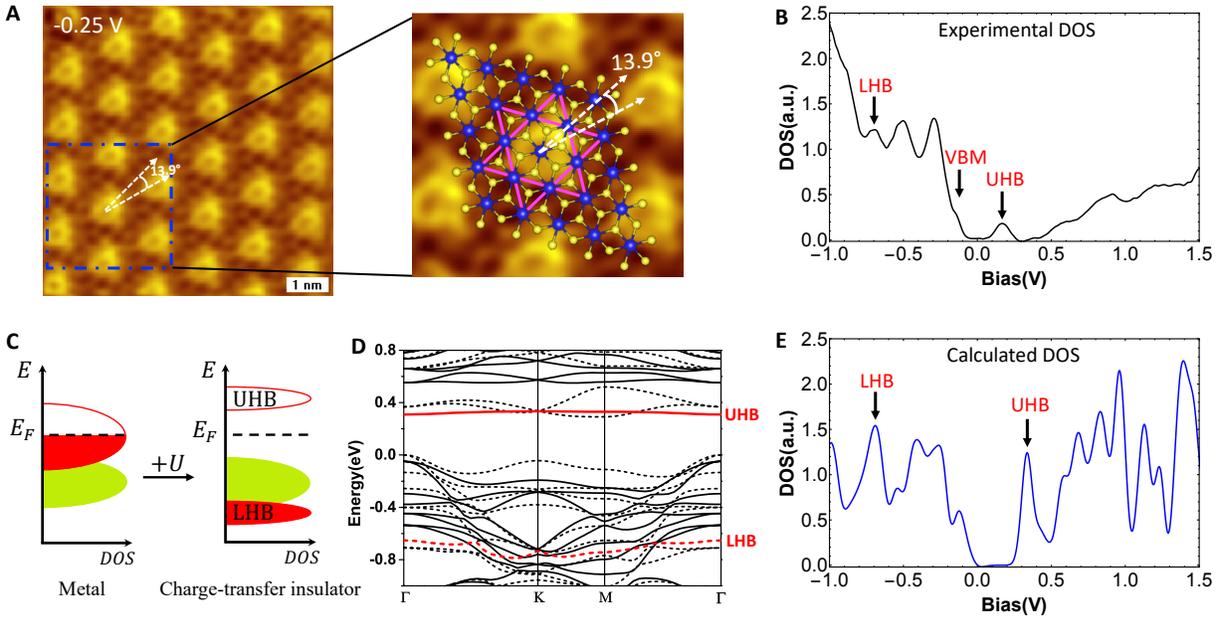

**Fig. 2. Charge-transfer insulator nature of 1T-NbSe$_2$.** (**A**) Atomic resolution topographic image of 1T-NbSe$_2$. The area enclosed by the blue dashed square is enlarged with an enhanced color contrast. The 1T-NbSe$_2$ lattice model superimposed on top is aligned with the atomic sites and the white dashed arrows show the rotation of the CDW relative to the lattice. (**B**) The d$I$/d$V$ STS spectrum of 1T-NbSe$_2$. The UHB is above the insulating gap and relatively isolated from the conduction band continuum; the LHB is below the VBM and merged with the valence band continuum, as marked by the black arrows. (**C**) Schematic illustration of energy levels for charge-transfer insulator driven by electron correlation U. (**D**) Calculated band structure of 1T-NbSe$_2$ in the $\sqrt{13} \times \sqrt{13}$ CDW ground state. The dashed and solid lines represent up and down spin bands, respectively. The UHB and LHB are highlighted in red. (**E**) Calculated DOS. The black arrows indicate the Γ point energy of UHB and LHB.

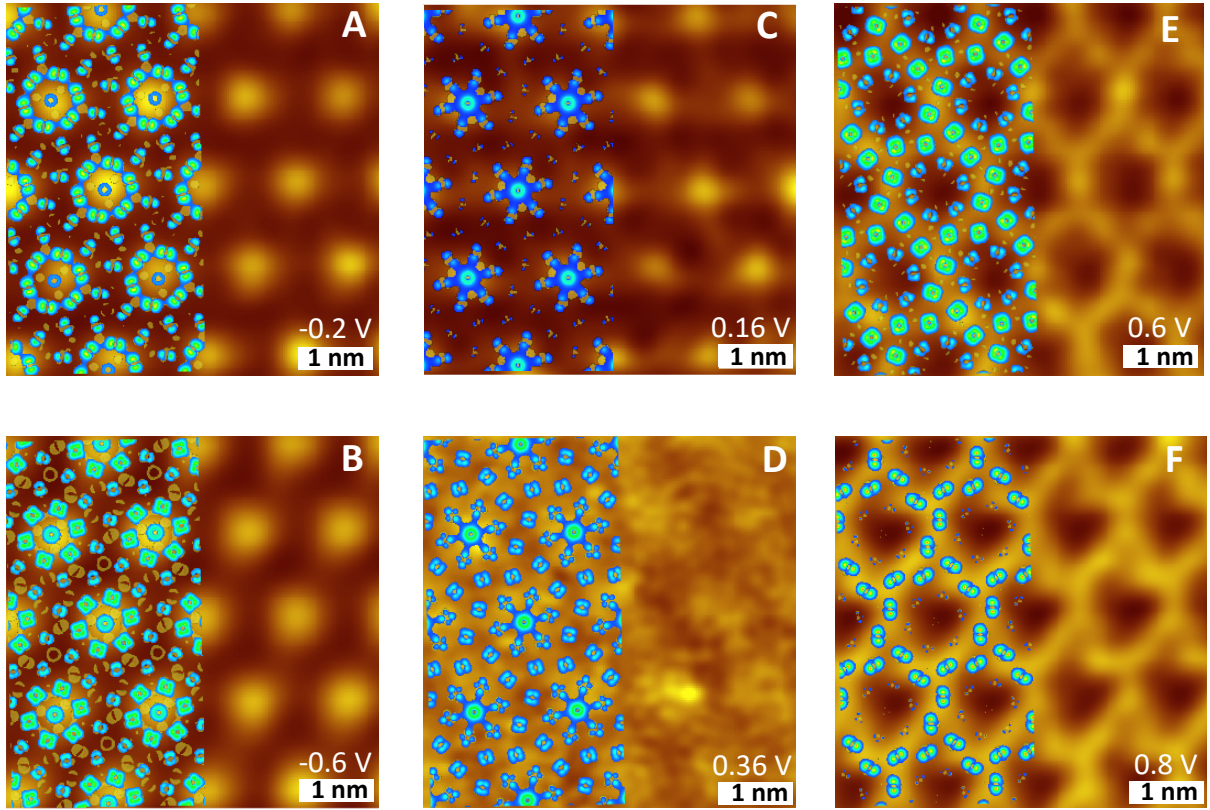

**Fig. 3. Orbital texture mapping of 1T-NbSe$_2$.** **(A to F)** Constant-height d$I$/d$V$ images of 1T-NbSe$_2$ at different biases showing energy dependent orbital textures. The calculated charge density distributions are overlaid on the left half of each d$I$/d$V$ image. Energies at which calculations are performed: (A): $E_{VBM}$- 0.25 eV; (B): $E_{VBM}$-0.65 eV; (C): $E_{UHB}$; (D): $E_{UHB}$+ 0.25 eV; (E): $E_{UHB}$+ 0.45 eV; (F): $E_{UHB}$+ 0.7 eV. $E_{VBM}$ and $E_{UHB}$ refer to the energies of VBM and UHB, respectively.

Page **17** of **20**

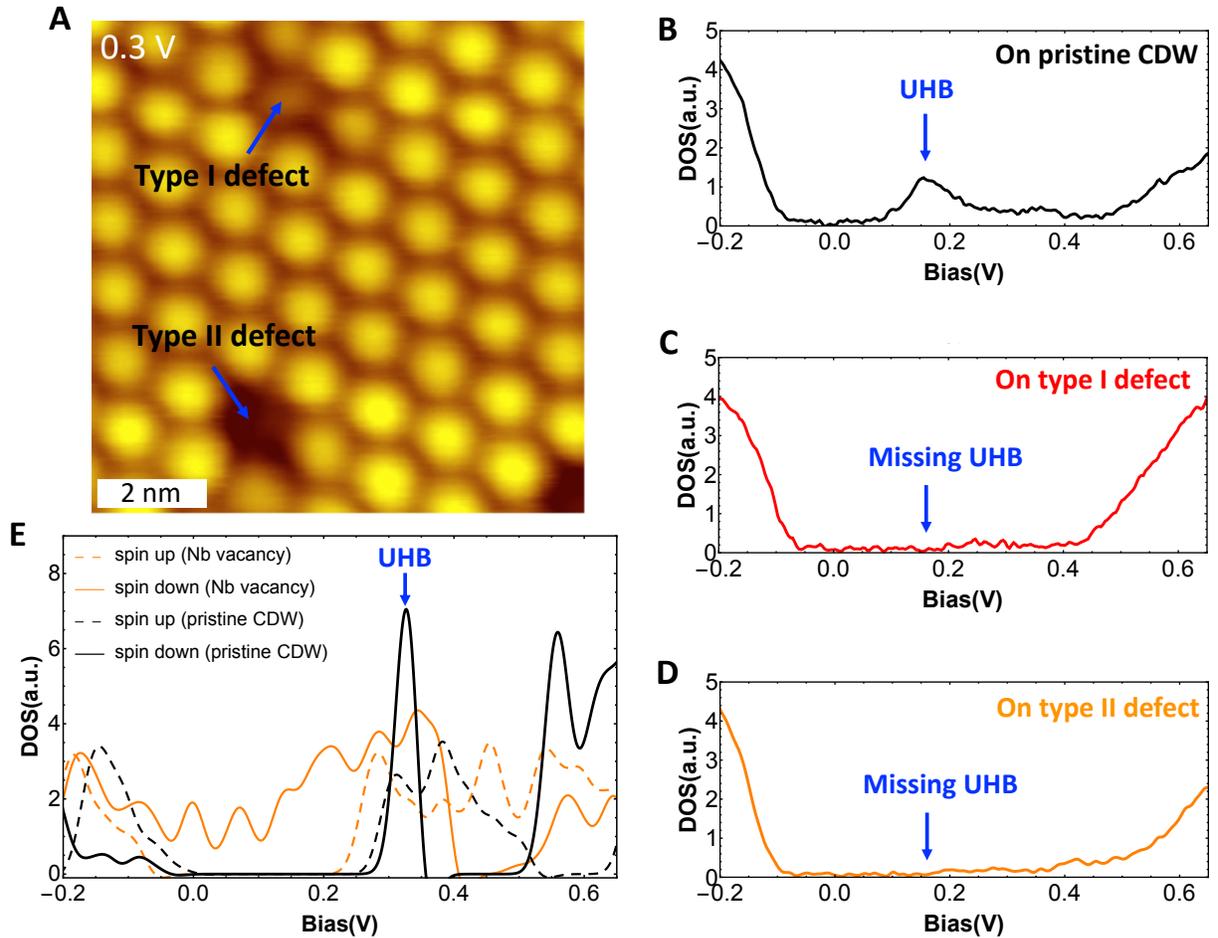

**Fig. 4. The fragility of UHB at local defect positions. (A)**, Two types of common defects in 1T-NbSe$_2$. Type I defects label weakened CDW spot. Type II defects label missing CDW spot. (**B to D**), Typical d$I$/d$V$ spectra taken at pristine CDW, type I and type II defects. In contrast to pristine CDW, d$I$/d$V$ spectra on type I and type II defects show missing UHB. (**E**), Calculated DOS for Nb vacancy model (orange lines) and pristine CDW (black lines). The dash and solid lines represent the up and down spin bands, respectively. The spin polarized UHB appears in pristine CDW but is missing in Nb vacancy model.

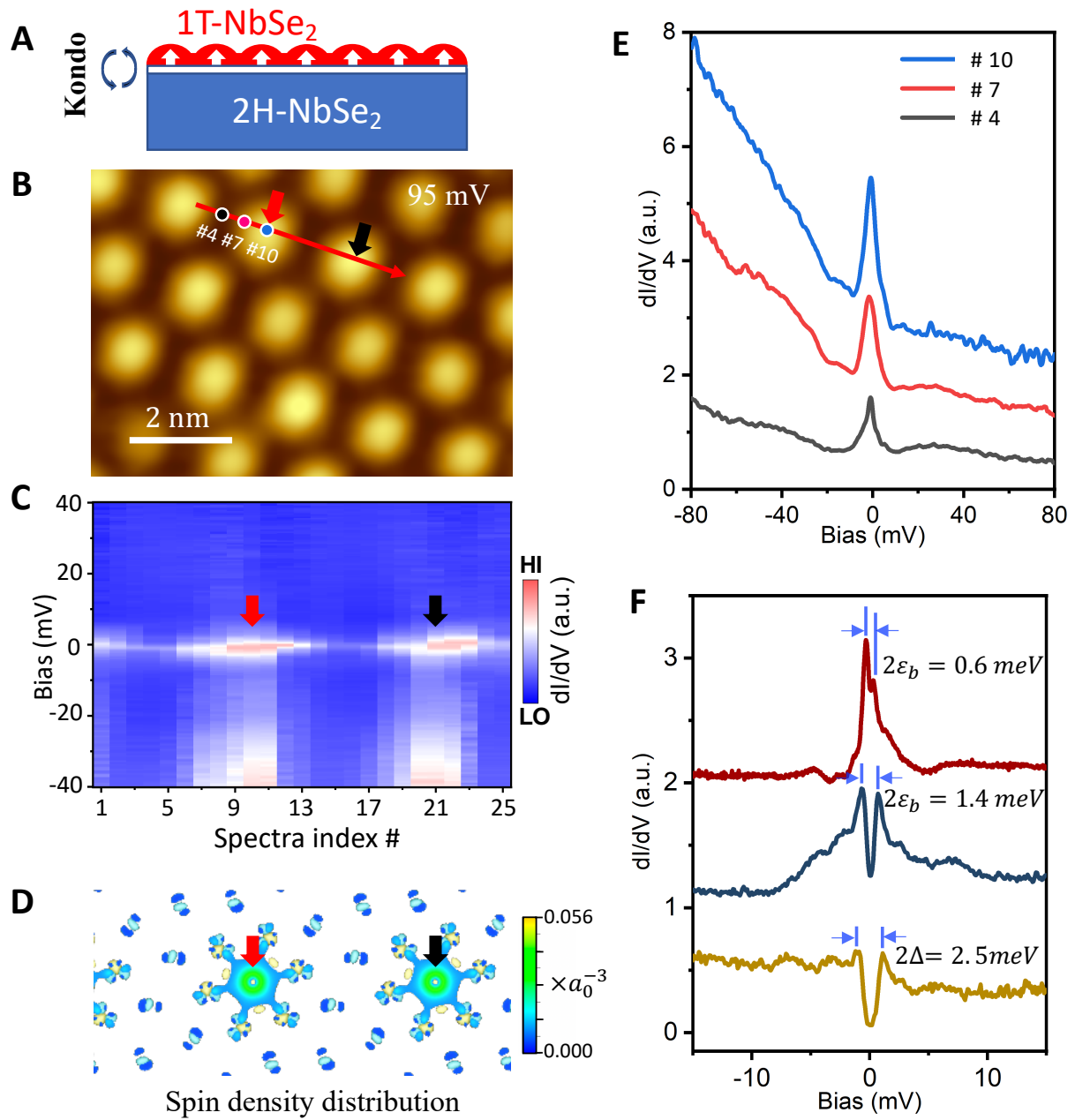

**Fig. 5. Localized magnetic moment in 1T-NbSe$_2$.** (**A**) Schematics of vertical heterostructure formed by monolayer 1T-NbSe$_2$ on bulk 2H-NbSe$_2$. The height profile in red represents the 1T-NbSe$_2$ CDW. Each unit cell carries a net magnetic moment illustrated by a white arrow and their directions are artificially set to be parallel. The coupling between the conduction electrons in 2H-NbSe$_2$ and the magnetic moment in 1T-NbSe$_2$ was manifested by the Kondo effect. (**B**) STM image on the heterostructure surface showing the CDW of 1T-NbSe$_2$. (**C**) False-color images of

d$I$/d$V$ spectra taken along the red line in (B). A periodic modulation of the Kondo peak amplitude was observed, and the two maxima marked by the red and black arrows correspond to the CDW centers. (**D**) Calculated spin density distribution in unit of $a_0^{-3}$, where $a_0$ is the Bohr radius. The red and black arrows mark the CDW centers. (**E**) Selected d$I$/d$V$ spectra from (C) (index number #4, #7, #10). Their positions are marked in (B) with the corresponding colors. (**F**) d$I$/d$V$ spectra with finer energy resolution on the heterostructure surface demonstrate the coexistence of YSR bound state and the Kondo resonance (red and dark blue lines); $\varepsilon_b$ labels the binding energy of the YSR states. For comparison, the dark yellow line plots the d$I$/d$V$ spectrum taken on pristine bulk 2H-NbSe$_2$ which exhibits superconductivity; $\Delta$ labels the superconducting gap.

Supplementary Materials for

# Monolayer 1T-NbSe$_2$ as a 2D correlated magnetic insulator

Mengke Liu[1], Joshua Leveillee[1,2], Shuangzan Lu[3], Jia Yu[1], Hyunsue Kim[1], Keji Lai[1], Chendong Zhang[3], Feliciano Giustino[1,2], Chih-Kang Shih[1]*
*email: shih@physics.utexas.edu.

**This PDF file includes:**

Supplementary Text
Fig. S1. Monolayer 1T and 1H phases of NbSe$_2$ on HOPG and graphene substrates.
Fig. S2. STS at local defect positions and Nb vacancy model.
Fig. S3. The creation of 1T/2H NbSe2 heterostructure.

**Supplementary Text**

**Monolayer 1T and 1H phases of NbSe$_2$ on HOPG and graphene substrate**

A single 1T phase NbSe$_2$ can be achieved on HOPG substrate. Fig. S1A shows the large-scale STM topographic image. A zoomed-in view of the flake is shown in Fig. S1B. The nominal flakes sizes are about 100 nm and the $\sqrt{13} \times \sqrt{13}$ charge density wave can be imaged at relatively high bias. Monolayer 1T-NbSe$_2$ can also be achieved on the graphene substrate. Bilayer graphene terminated 6H-SiC(0001) substrate is used and Fig. S1C shows its typical atomic image. However, on the graphene substrate (Fig. S1D), 1H phase flakes usually coexist with 1T phase flakes. In direct contrast to the 1T flakes whose CDW corrugations are clearly visualized at relatively high bias, the 1H flakes appear to be uniform and flat.

**STS at local defects position and Nb vacancy model**

Fig. S2A shows a zoomed-in view of the type I and type II defects. With the enhanced color contrast of the STM image, the distinct CDW morphology between type I and type II defect are clearly visualized. Fig. S2B shows the Nb vacancy model we used in our DFT calculation. The central Nb atom in each David star cell is removed.

**The creation of 1T/2H NbSe$_2$ heterostructure**

A 2H to 1T phase transformation on bulk NbSe$_2$ surface can be induced by applying an electric pulse using the STM tip, as schematically illustrated in Fig. S3A. Fig. S3B shows the resulted STM topographic image. The coexistence of 1T and 2H NbSe$_2$ is visualized in Fig. S3B.

# Figures

**Fig. S1.**

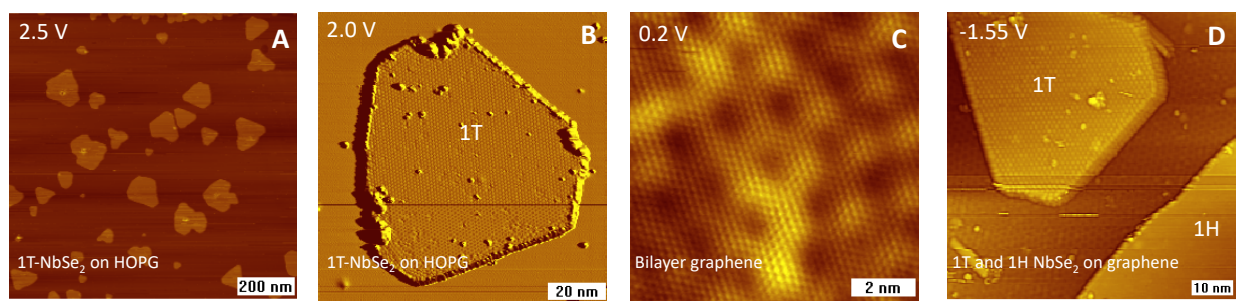

**Fig. S1. Monolayer 1T and 1H phases of NbSe$_2$ on HOPG and graphene substrates. (A)** Large-scale STM topographic image of single 1T phase NbSe$_2$ on HOPG substrate. **(B)** A zoomed-in view of 1T-NbSe$_2$ flakes on HOPG substrate. STM image is in light shade mode for clear recognition of the $\sqrt{13} \times \sqrt{13}$ CDW corrugation. **(C)** Atomic image of bilayer graphene/SiC(0001) substrate. **(D)** Mixed 1T and 1H phases of NbSe$_2$ on graphene substrate.

**Fig. S2.**

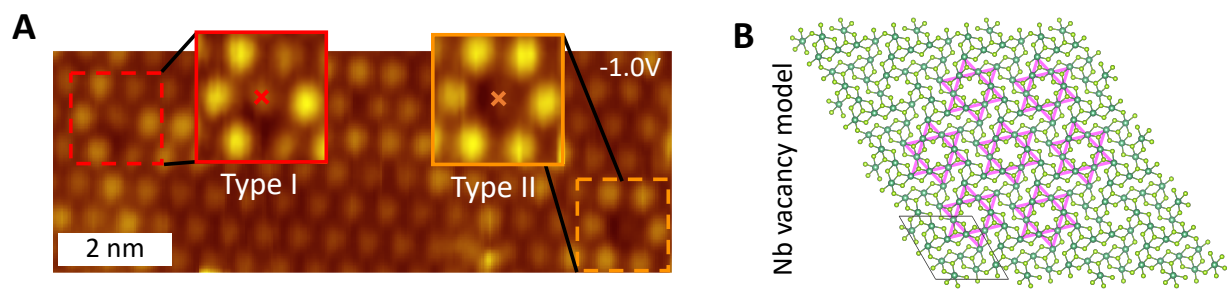

**Fig. S2. STS at local defect positions and Nb vacancy model. (A)** STM image of type I and type II defects. The dashed square enclosed areas are enlarged in the solid square with an enhanced color contrast for recognition of the CDW morphology. The red cross and orange cross mark the positions where STS for type I and type II defects (Fig. 4c and Fig. 4d) are taken. **(B)** Nb vacancy model used in DFT calculation. A few David star supercells are outlined in magenta.

**Fig. S3.**

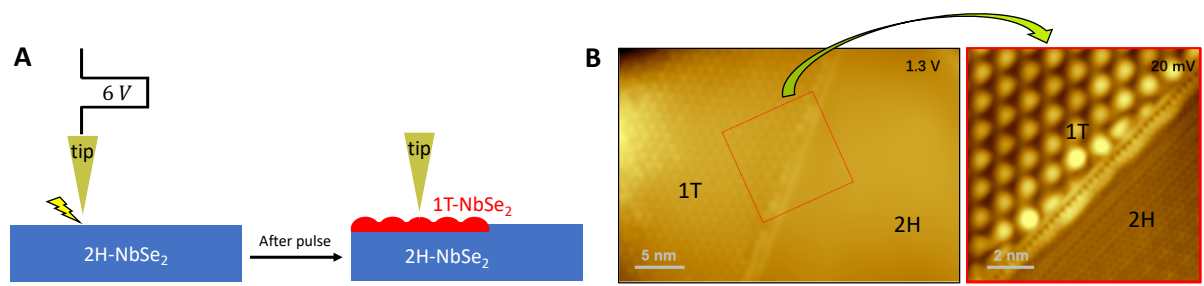

**Fig. S3. The creation of 1T/2H NbSe$_2$ heterostructure. (A)** Schematic illustration of creating monolayer 1T-NbSe$_2$ on bulk 2H-NbSe$_2$ surface. A 6 V pulse applied from the STM tip to the 2H-NbSe$_2$ induces a monolayer 1T-NbSe$_2$ on the surface. **(B)** STM topographic image showing the coexistence of 1T and 2H NbSe$_2$. The red dashed square enclosed area is enlarged in the solid red square and with an enhanced color contrast for clear recognition of the 1T and 2H phase.